# Real-Time Super-Resolution Imaging System Based on Zero-Shot Learning for Infrared Non-Destructive Testing

Pengfei Zhu, Student Member, IEEE, Ziang Wei, Ahmad Osman, Clemente Ibarra-Castanedo, Andreas Mandelis*, Xavier Maldague, Senior Member, IEEE, and Hai Zhang*, Member, IEEE

*Abstract*—Infrared thermography (IRT) and photothermal coherence tomography (PCT) exhibit potential in non-destructive testing and biomedical fields. However, the inevitable heat diffusion significantly affects the sensitivity and resolution of IRT and PCT. Conventional image processing techniques rely on capturing complete thermal sequences, which limits their ability to achieve real-time processes. Here, we construct a real-time super-resolution imaging system based on zero-shot learning strategy for the non-invasive infrared thermography and photothermal coherence tomography techniques. To validate the feasibility and accuracy of this super-resolution imaging system, IRT systems were employed to test several industrial samples and one biomedical sample. The results demonstrated high contrast in the region of interest (ROI) and uncovered valuable information otherwise obscured by thermal diffusion. Furthermore, three-dimensional photothermal coherence tomography was used to validate the excellent denoising and deconvolution capabilities of the proposed real-time super-resolution imaging system.

## I. INTRODUCTION

INFRARED thermography (IRT) is an attractive non-destructive testing (NDT) method that provides rapid, full-field, non-contact inspection of materials [1], [2]. Based on its excitation modalities, IRT can be divided into pulsed thermography (PT), line scanning thermography (LST), modulated thermography (MT), etc. [3], [4]. However, conventional OET techniques offer relatively limited two-dimensional subsurface structure information.

In recent years, photothermal coherence tomography (PCT) [5] techniques have been intensively developed to provide air-coupled, fast, optical-to-thermal energy conversion modalities for molecular, spectroscopic, and nondestructive [6], [7] imaging applications using non-scanning multi-array infrared (IR) cameras coupled with ultrafast (speed-of-light) transmission of thermal IR photons (radiative emission channel). For instance, the truncated correlation photothermal coherence thermography (TC-PCT) technique [8] not only provides depth-resolved tomographic images and three-dimensional (3D) mapping of biological hard and soft tissues but also overcomes the issues of low resolution and inhomogeneous thermal perturbations found in conventional dynamic thermal tomography techniques [9]. Nevertheless, the diffusion of thermally converted optical energy spreads and increases the effective size of features from internal structure, leading to a reduction in spatial resolution and contrast proportional to the depth of the feature below the surface.

To mitigate the thermal diffusion effect in infrared thermography, many effective signal and image processing methods have been proposed. In 1996, Maldague *et al.* [10] introduced a pulse phase thermography (PPT) method based on the Fourier transformation to extract valuable low-frequency

This work was supported in part by the Natural Sciences and Engineering Research Council of Canada (NSERC) through the CREATE-oN DuTy! Program under Grant 496439-2017, in part by the Discovery Grants Program under Grant RGPIN-2020-04595, in part by the Canada Research Chair in Multi-polar Infrared Vision (MIVIM), in part by the Canada Foundation for Innovation (CFI) Research Chairs Program under Grant 950-230876, and in part by the CFI-JELF program (38794). (Corresponding authors: Andreas Mandelis; Hai Zhang.)

Pengfei Zhu, Ziang Wei, Clemente Ibarra-Castanedo, Xavier Maldague and Hai Zhang are with the Department of Electrical and Computer Engineering, Computer Vision and Systems Laboratory (CVSL), Laval University, Québec G1V 0A6, Québec city, Canada (e-mail: pengfei.zhu.1@ulaval.ca; wei.ziang.1@ulaval.ca; clemente.ibarra-castanedo@gel.ulaval.ca; xavier.maldague@gel.ulaval.ca; hai.zhang.1@ulaval.ca). Hai Zhang is also with the Centre for Composite Materials and Structures (CCMS), Harbin Institute of Technology, Harbin 150001, China (hai.zhang@hit.edu.cn).

Ahmad Osman is with the Saarland University of Applied Sciences and Fraunhofer Institute for Nondestructive Testing, Germany (e-mail: ahmad.osman@izfp.fraunhofer.de).

Andreas Mandelis is with the Department of Mechanical and Industrial Engineering, Center for Advanced Diffusion-Wave and Photoacoustic Technologies (CADIPT), University of Toronto, Toronto, ON M5S 3G8, Canada, and also with the Department of Mechanical and Industrial Engineering, Institute for Advanced Non-Destructive and Non-Invasive Diagnostic Technologies (IANDIT), University of Toronto, Toronto, ON M5S 3G8, Canada (email: mandelis@mie.utoronto.ca).

information. Notably, phase images can effectively reduce the thermal diffusion effect and increase both axial and lateral resolution. Shepard *et al*. [11] proposed a thermal signal reconstruction (TSR) algorithm, which relies on the linear relationship between temperature and time in the logarithmic domain. Rajic developed principal component analysis (PCoT) to detect defects in composite materials [12]. PCoT relies on principal component analysis (PCA), which is based on singular value decomposition (SVD) for feature extraction. Subsequently, Yousefi *et al*. [13] proposed the application of candid covariance-free incremental principal component analysis (CCIPCA) in thermography, which demonstrated computational efficiency and an incremental, covariance free version of the original PCA method. Lopez *et al.* [14] applied a statistical correlation method, partial least squares regression (PLSR), to experimental PT data from a carbon fiber-reinforced composite with simulated defects, showing that PLSR has a similar effect to PCA. Zhang *et al*. [15] employed the PCA method in photothermal coherence tomography, effectively improving the image quality (defect contrast), though the original tomograms lost depth features. Recently, Thapa *et al.* [16] proposed a spatiotemporal gradient adaptive filtering to increase spatial resolution based on Richardson-Lucy deconvolution [17]. The frame difference method was introduced by Zhu et al. [18] which, however, results in the rapid deterioration of 3D reconstruction results, similar to Zhang's report.

The aforementioned signal / image processing methods are based on principles from statistics or frequency analysis. However, these methods offer limited improvement for detecting deeper defects/structures and relying significantly on parameter sets. With the advancement of artificial intelligence, deep learning networks have begun to play an important role in infrared thermography [19]. For instance, Yolo series networks have been used for automatic defect detection [20]. U-Net networks have been used for damage segmentation [21]. Generative adversarial network for super-resolution (SR) imaging [22]. However, these deep learning-based methods (DLBM) require extensive datasets and ground truth (GT) images which are challenging to obtain in infrared thermography. Zero-shot learning [23], as a powerful tool without requiring any prior image examples or prior training, can be explored to solve the data hungry issue in non-destructive testing (NDT) fields.

Here, we construct a real-time super-resolution imaging system which processes single frame image from infrared camera. It can effectively mitigate the heat diffusion effect, based on a zero-shot deep learning network interacting with a specific processing modality (using multi-scale outputs from one network which replaces two progressive networks for denoising and deconvolution), as shown in Fig. 1. The image recorrupting scheme (where the original clean image is corrupted twice using different noise patterns) proposed in [24] is used to generate two noise-independent recorrupted images from the original image as the input image and target image. The deep neural network (DNN) U-Net++ [25] is employed as the backbone, with two different outputs defined from two feature layers. The first output is used for denoising, and the second output is convolved with the point spread function (PSF) for deconvolution. Several industrial specimens, including glass fiber reinforced polymer (GFRP) laminates and plexiglass plates, as well as biomedical specimens were used to validate the feasibility of the proposed method. The results demonstrated excellent denoising capability in both 2D and 3D imaging.

II. ZERO-SHOT MULTI-SCALE NEURAL NETWORKS

As shown in Fig. 1(a), real-time super-resolution imaging system consists of six steps: image capturing, access to PySpin, load model, format conversion, visualization, save images. Of note, the embedded network cannot be a large-scale model, as this would significantly increase inference time and introduce delays. Before introducing the proposed network, we present two alternatives: one is a state-of-the-art zero-shot network, and the other is a classic super-resolution algorithm.

*A. Zero-Shot Deconvolution Networks*

Noise in IR images is mainly additive. For a noisy image $y = x + n$ where $x$ is the noise-free counterpart, $n$ is the random noise and follows the normal distribution $\mathcal{N}(0, \Sigma_x)$. Typical supervised learning methods train the deep neural networks (DNNs) using

$$\min_\theta \mathbb{E}_{x,y} \mathcal{L}(f_\theta(y), x) \tag{1}$$

where $\mathcal{L}(\cdot,\cdot)$ denotes the loss function, $f_\theta$ is a DNN with trainable parameters $\theta$, and $\mathbb{E}_{x,y}$ is the expectation over the joint distribution of the clean and noisy image pairs ($x$, $y$). If there is no access to noise-free images, the objective function above can be re-written as

$$\min_\theta \mathbb{E}_y \|f_\theta(y) - y\|_2^2 \tag{2}$$

where $\|\cdot\|_2^2$ denotes the squared $\ell_2$-norm loss. In this case, the DNN does not remove any noise but outputs the noisy image itself. The training Recorrupted-to-Recorrupted (R2R) scheme [24] is employed in this work to generate paired images $\{(\hat{y}, \tilde{y})\}$, where $\hat{y} = y + D^T n$, $\tilde{y} = y - D^{-1} n$, and $n \sim \mathcal{N}(0, \sigma^2 I)$. $D$ is an invertible unit matrix. $\sigma^2 = H(y - b)$ is the variance of this Gaussian distribution noise map, where $b$ is the background (the original signal before heating). $H(\cdot)$ is a linear low-pass filter (averaging filter with a size of 5 pixels) used to preliminarily smoothen the image and reduce the noise. For general optical imaging systems, the unsupervised inverse problem solver can be constructed as:

$$\mathrm{argmin}_\theta \|y - (f_\theta(y) * \mathrm{PSF})_\downarrow\|_2^2 \tag{3}$$

where PSF is the point spread function (PSF), and $(\cdot)_\downarrow$ is a down-sampling operation. If the DNN is trained directly via the above objective function, it will undesirably amplify the photon noise, which will substantially contaminate the real sample information at low signal-to-noise ratio (SNR) conditions. To avoid amplifying the photon noise, Qiao *et al.* [26] proposed the use of a ZS-DeconvNet, which designs two U-Net networks, one for denoising and the other for deconvolution. The loss function can then be given as:

$$\mathcal{L}(\hat{y}, \tilde{y}) = \mu \mathcal{L}_{den}(\hat{y}, \tilde{y}) + (1-\mu)\mathcal{L}_{dec}(\hat{y}, \tilde{y}) \quad (4a)$$

$$\mathcal{L}_{den}(\hat{y}, \tilde{y}) = \|f_{\theta'}(\hat{y}) - \tilde{y}\|_2^2 \quad (4b)$$

$$\mathcal{L}_{dec}(\hat{y}, \tilde{y}) = \|(f_\theta(\hat{y}) * \text{PSF})_\downarrow - \tilde{y}\|_2^2 + \lambda \mathcal{R}_{Hessian}(f_\theta(\hat{y})) \quad (4c)$$

where $\mu$ and $\lambda$ are scalar weighting factors, $f_{\theta'}(\hat{y})$ and $f_\theta(\hat{y})$ are the output images of the denoising stage ($\mathcal{L}_{den}(\hat{y}, \tilde{y})$) and the deconvolution stage ($\mathcal{L}_{dec}(\hat{y}, \tilde{y})$). To mitigate reconstruction artifacts and regulate the network convergence, the Hessian regularization term $\mathcal{R}_{Hessian}$ is used. As is well-known, introducing two networks significantly increases the overall complexity of the model, thereby requiring more computational resources and training time. The quality of the output of a denoising network directly affects the quality of the input for a deconvolution network. In addition, hyperparameter tuning and overfitting considerably increase the difficulty of debugging and optimization.

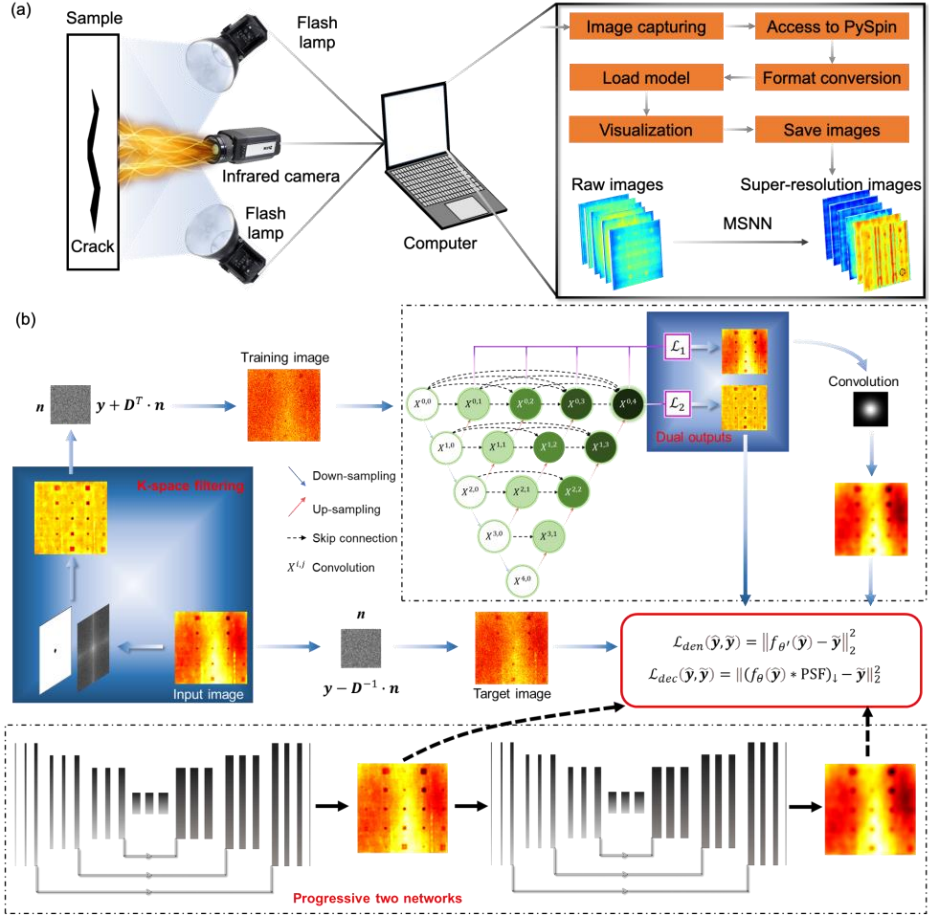

**Fig. 1.** The constructed real-time super-resolution system: (a) schematic image. (b) architecture of the embedded zero-shot multi-scale neural network (ZS-MSNN)

*B. Richardson-Lucy Deconvolution*

The RL method [27], similar to naive inverse filtering, follows a maximum-likelihood approach. However, unlike inverse filtering, the RL algorithm assumes that the noise follows a Poisson distribution, which results in

$$I^{t+1} = I^t \times \tilde{k} * \frac{B}{k \otimes I^t} \quad (5)$$

where $B$ is the observed blurred image, $k$ is the point spread function (PSF), $I$ is the clear image desired to estimate, $\tilde{k}$ is the transpose of $k$ that flips the shape of $k$ upside-down and left-to-right, $*$ is a convolution operation, $\times$ is a pixel-wise multiplication operation and $t$ is the number of iterations.

As a maximum-likelihood algorithm, the RL method is susceptible to the same noise amplification issue. Therefore, the optimal number of iterations should be determined heuristically to halt the algorithm before full convergence. To address this, Dey et al. [27] introduced total variation (TV) regularization, which mitigates noise amplification during deconvolution by minimizing the gradient magnitude in the blurred image

$$R_{TV}(I) = \int \sqrt{\|\nabla I(x)\|^2} dx \quad (6)$$

where $\nabla I(x)$ is the first order vector derivative of $I(x)$ (in the $x$ and $y$ directions). Substituting this regularization term into Eq. (1), one obtains

$$I^{t+1} = \left(\frac{B}{k \otimes I^t} * \tilde{k}\right) \times \frac{I^t}{1 - \xi \nabla R_{TV}(I)} \quad (7)$$

where $\nabla R_{TV}(I) = -\nabla \cdot (\frac{\nabla I^t}{|\nabla I^t|})$. It is important to note that this regularization may lead to division by zero or negative values, therefore, the regularization parameter $\xi$ should not be excessively large. Dey *et al.* recommended setting $\xi = 0.002$ and applied a specific convergence criterion

$$\frac{\sum_x \sum_y |I^{t+1}(x,y) - I^t(x,y)|}{\sum_x \sum_y I^t(x,y)} < \Lambda \tag{8}$$

where $\Lambda$ is the pre-determined threshold.

A theoretical PSF is calculated from the Born-Wolf (BW) model [28]

$$\kappa_\theta^{BW}(x) = C_\theta^{BW} | \int_0^1 J_0(k_0 r \text{NA} \rho) e^{-i\Phi(\rho,z)} \rho d\rho |^2 \tag{9}$$

where $C_\theta^{BW}$ is the normalization constant, $J_0$ denotes the Bessel function of the first kind of order zero, $k_0 = \frac{2\pi}{\lambda n_i}$ is the angular wave number in vacuum, $\Phi(\rho, z) = \frac{k_0 z \text{NA}^2 \rho^2}{2 n_i}$ is the phase term, $r = \sqrt{x^2 + y^2}$, and NA is the numerical aperture. The model is parameterized by three parameters, *i.e.*, the emission wavelength, the numerical aperture, and the refractive index of the immersion medium, denoted by $\theta^{BW}(x) = \{\lambda, \text{NA}, n_i\} \in \Theta^{BW} = \mathbb{R}_+^3$.

*C. Zero-Shot Multi-Scale Neural Network*

The core unit of the real-time super-resolution system is a zero-shot multi-scale neural network (ZS-MSNN), which uses a dual output scheme for replacing the denoising and deconvolution capabilities from two progressive networks. To reduce the model complexity from two progressive networks, U-Net++ [23] was found to be a good choice as the backbone of this scheme because U-Net++ contains multi-branches with different dimensional features, as shown in Fig. 1. In U-Net++, the feature maps of the encoder undergo a dense convolution block the number of convolution layers of which depends on the pyramid level. The stack of feature maps represented by $x^{i,j}$ is computed as

$$x^{i,j} = \begin{cases} \mathcal{H}(x^{i-1,j}), & j = 0 \\ \mathcal{H}\left([x^{i,k}]_{k=0}^{j-1}, u(x^{i+1,j-1})\right), & j > 0 \end{cases} \tag{10}$$

where the function $\mathcal{H}(\cdot)$ is a convolution operation followed by an activation function; $u(\cdot)$ denotes an up-sampling layer; $[\cdot]$ denotes the concatenation layer; $i$ indexes the down-sampling layer along the encoder and $j$ indexes the convolution layer of the dense block along the skip pathway. It should be noted that there are two outputs for the U-Net++ network ($\mathcal{L}_1$ and $\mathcal{L}_2$). $\mathcal{L}_1$ is an accurate mode wherein the outputs from all branches are averaged, and $\mathcal{L}_2$ is a fast mode wherein the final segmentation map is selected from only the final branch $X_{0,4}$. Then the accurate mode $\mathcal{L}_1$ undergoes a convolution operation with the PSF function. The loss function can be given as:

$$\mathcal{L}(\hat{y}, \tilde{y}) = \mu \| f_{\theta'}(\hat{y}) - \tilde{y} \|_2^2 + (1 - \mu) \| (f_\theta(\hat{y}) * \text{PSF})_\downarrow - \tilde{y} \|_2^2 \tag{11}$$

The hyperparameter $\mu$ was empirically set to 0.5. It was validated to be stable on all the samples for a large $\mu$ range [26]. In addition, since interference in infrared thermography predominantly arises from low frequency noise, the interference in this work was selected from the low frequency noise. A 2D Fourier high-pass (k-space) filter with a constant frequency component of 5 was empirically employed for pre-processing the input images according to preliminary tests.

III. EXPERIMENTS AND TRAINING SETUPS

Pulsed infrared thermography offers advantages of easy implementation, broad scope, and high efficiency, making it suitable for industrial inspections for many years. It can be described using 1D analytical models [29], assuming that a Dirac delta pulse with energy $q_0$ heats the surface. The surface temperature $T$, as a function of time, is given by [29]:

$$T(0, t) = \frac{q_0}{\rho C_p L} [1 + 2 \sum_{n=1}^\infty \exp(-\frac{n^2 \pi^2 \alpha t}{L^2})] \tag{12}$$

where $\alpha = k/(\rho C_p)$ is the thermal diffusivity, $C_p$ is the specific heat at constant pressure, $\rho$ is the density of the material, $k$ is the thermal conductivity, and $L$ is the thickness of the sample. Fig. 2 shows the PT experimental setup in the reflection mode. A cooled infrared camera (FLIR X8501sc, 3-5 μm, InSb, NEdT < 20 mK, 1280 × 1024 pixels) and two Xenon flashes (Balcar, 6.4 kJ for each, 2 ms) were used for photothermal imaging.

In general, zero-shot learning does not require any datasets. However, to make the proposed network suitable for processing infrared thermography images, we selected open-source datasets from [25] without labels to train our model. All experimental results in the following sections were based on training datasets from Ref. [25]. The training process was conducted using a UNet++-based model for image denoising and deconvolution, following an unsupervised learning approach with an Adam optimizer. The network was trained for 50 epochs with a batch size of 16, an initial learning rate of 1×10$^{-4}$ and a learning rate decay factor of 10 applied after 30 epochs. Training and validation images were corrupted with Gaussian noise at a standard deviation of 25. Two training strategies were used: Noise2Clean (N2C), where the clean image served as the target, and Recorrupted-to-Recorrupted (R2R), where additional low-frequency noise perturbations were applied to create input-output pairs. The model was initialized using Kaiming initialization [30] and trained on a single NVIDIA 4060 Titan GPU (CUDA). At each epoch, the network's performance was evaluated on a separate validation dataset using Peak Signal-to-Noise Ratio (PSNR) [31] as the primary metric. Loss and PSNR values were logged, and learning rate adjustments were applied as needed. The final model weights were saved, and training progress was visualized through loss and PSNR curves to assess model convergence. The overall training time is ~2 h. The model size is only 45.9 MB (48,219,557 bytes). The training parameters

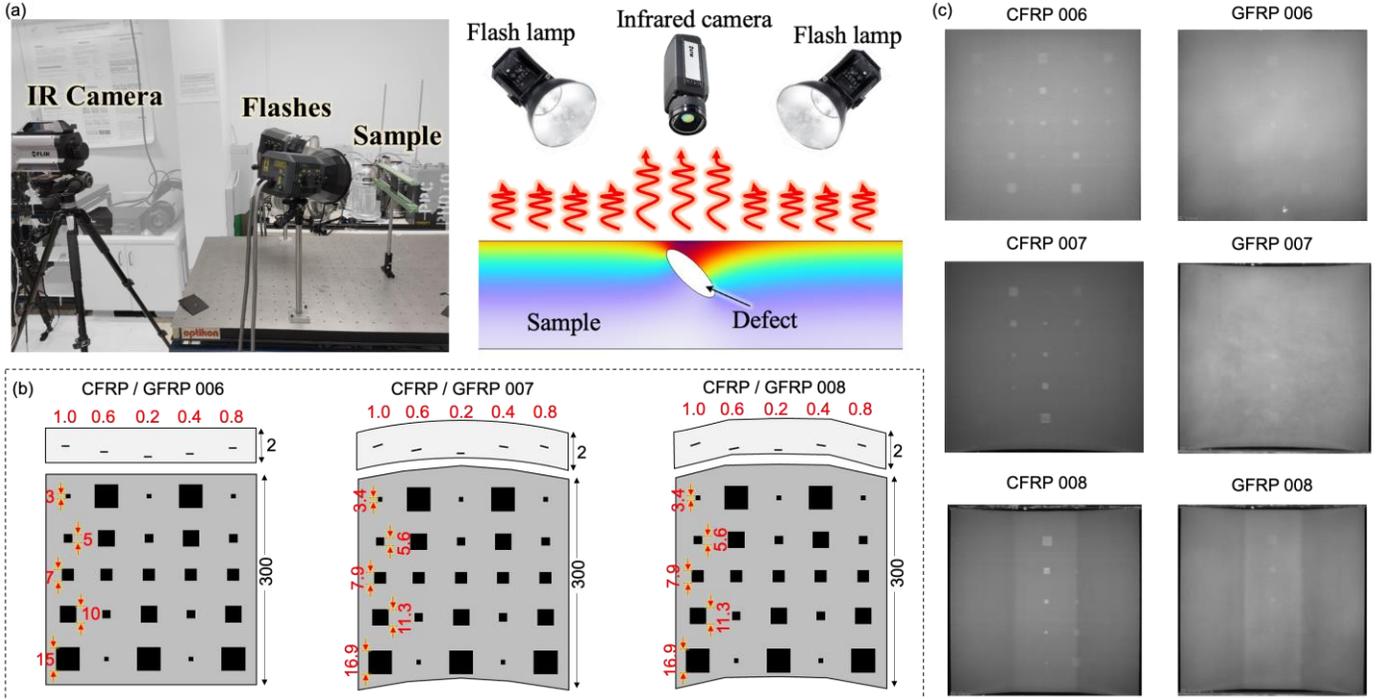

Fig. 2. Experimental setup of pulsed thermography and training dataset of ZS-MSNN: (a) Experimental setup, (b) Schematic image of six plates, (c) Original results of six plates.

are shown in Table I.

TABLE I
TRAINING PARAMETERS.

| Parameter | Value |
|---|---|
| Batch size | 16 |
| Epochs | 50 |
| Initial learning rate | $1\times10^{-4}$ |
| Optimizer | Adam |
| Training noise level | 25 |
| Training strategy | R2R / N2C |
| Weight initialization | Kaiming Initialization |

IV. RESULTS AND DISCUSSION

*A. Training Performance*

It is noted that the training process is performed before running infrared cameras. Because the real-time super-resolution imaging system is based on zero-shot learning, only a few shots (without labels) are required to train the embedded networks. In this work, the training datasets [32] consisted of six composite plates with the same dimensions (300 mm × 300 mm × 2 mm) and defect distributions but made from different materials: three plates were made of carbon fiber-reinforced plastic (CFRP) and another three plates of glass fiber reinforced plastic (GFRP) with three different geometries: planar, curved, and trapezoidal (Fig. 2(b)). The details of defect sizes/depths are described in Fig. 2(b). The original thermograms of the six plates are shown in Fig. 2(c).

To exhibit the powerful denoising capability, a pulse phase thermography algorithm (PPT) [10] was first employed to extract useful information for the defect location, as shown in Fig. 3(a). It was found that the background noise significantly reduces the detectability of defects. Furthermore, the uneven surface profile (curves in CFRP / GFRP 008) also generated unavoidable noise.

The peak signal-to-noise ratio (PSNR) was used to evaluate the network, using the following formula [31]

$$PSNR = 10 \log_{10} \frac{255^2}{M*N \sum_{i=1}^{M} \sum_{j=1}^{N} |R(i,j)-F(i,j)|^2} \quad (13)$$

where $F$, $R$, $M$, and $N$ represent the intensity of defect areas, intensity of sound areas, and image size.

As shown in Fig. 3(b), the loss function decreases with each epoch while PSNR increases, indicating positive trends in the training process. The training results are shown in Fig. 3(c). When the noise level $\chi$ is set to 25 (with standard deviation $\sigma = \chi / 255$), it is evident that ZS-MSNN effectively reduces noise and enhances defect contrast. The defects obscured by thermal diffusion are highlighted within the black circles. It should be noted that the ZS-MSNN approach differs from conventional signal processing algorithms in infrared thermography as it processes only a single image instead of the entire time-domain information. Therefore, it is not only real-time processing but can also be integrated with any other signal processing / 3D tomographic methods.

To quantify our results, the contrast-to-noise ratio (CNR) [33] and the full width at half maximum (FWHM) [34] of the testing results were calculated as the quantitative metrics. The CNR formula is given by [33]:

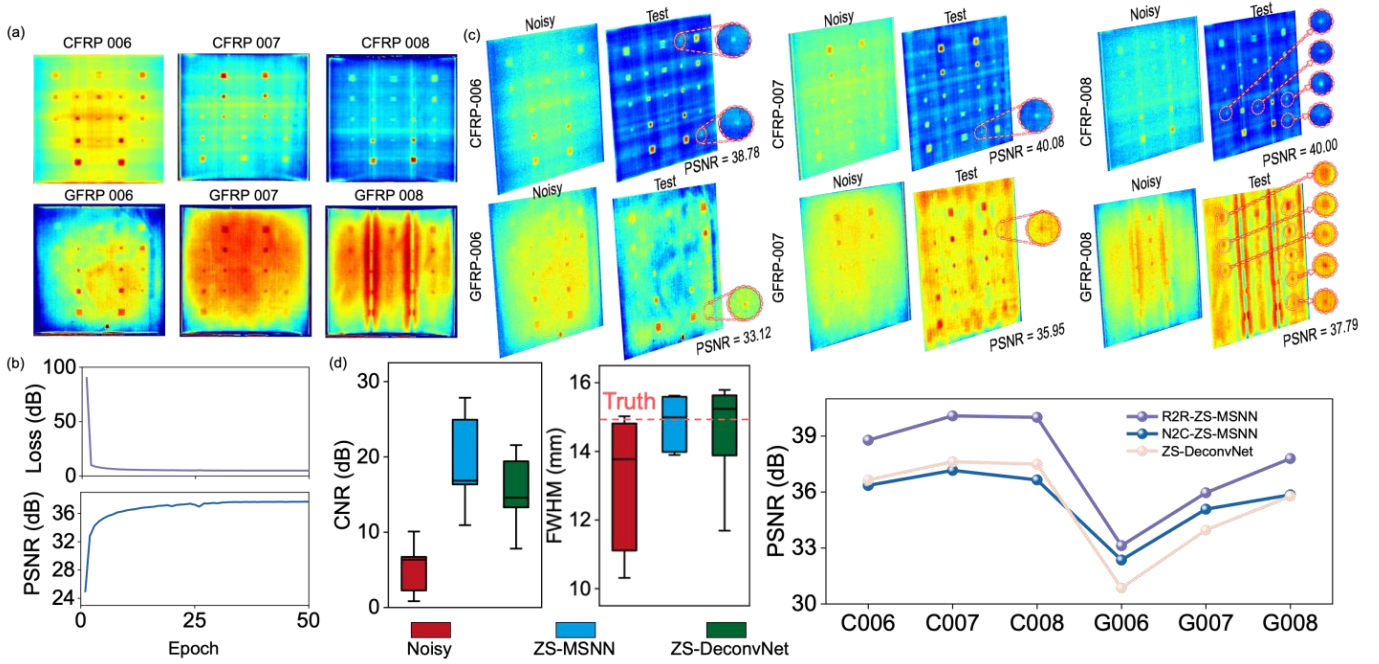

**Fig. 3.** The training results. (a) Tomograms after pulsed phase thermography (PPT) processing. (b) Loss function and PNSR during the training process. (c) Noised images (adding noise on PPT results) and denoising results (ZS-MSNN), where red circles denote the detected defects hidden due to thermal diffusion. (d) Quantitative evaluation of testing results. Top: the CNR (left) and FWHM (right) results from Noised, ZS-MSNN, and ZS-DeconvNet results. Bottom: The PSNR values used for evaluating three networks.

$$CNR = \frac{|\mu_d - \mu_s|}{\sigma_s} \qquad (14)$$

where $\mu_d$ and $\mu_s$ are, respectively, the mean values of the defective and the sound area, and $\sigma_s$ denotes the standard deviation of the sound area. The FWHM can be directly measured from the temperature profile of the defective area and its surrounding sound area.

The ZS-DeconvNet was used for comparison with our results (Fig. 3(d)). The results illustrated that ZS-MSNN achieves superior denoising and deconvolution performance. Furthermore, the general noise-to-clean (N2C) scheme [35] was employed to validate the effectiveness of the R2R scheme in infrared thermography. This demonstrates that the R2R scheme is not only suitable for denoising visual (visible range) images but also for infrared images. As is well-known, the noise in infrared images always co-exists with defect information in the low-frequency instead of the high-frequency range. Therefore, eliminating noise while retaining defect information is challenging. The ZS-MSNN provides a novel approach to address this issue.

### B. 2D Super-Resolution for Infrared Thermography

To further verify the generalization capability of the proposed model, we selected three plexiglass samples with different shape defects and one *ex vivo* mouse brain picture from Ref. [16]. In Fig. 4(a), it is obvious that uneven heating on both sides of plexiglass 1 significantly affects defect detection in the original image. The Richardson-Lucy deconvolution (RL-Deconv) algorithm fails to eliminate this type of noise and instead reduces the original defect contrast. Moreover, while ZS-DeconvNet effectively removes low frequency noise, it does not address the influence of uneven heating, and this type of noise continues to impact image quality due to thermal diffusion. The ZS-MSNN effectively removes the uneven heating noise and enhances defect contrast, as evidenced in Figs. 4(a), 4(b) and 4(d). In Figure 4(c), we chose the screenshot from Ref. [16]. The original image appears relatively blurred due to low spatial resolution resulting from the high frame rate in photothermal coherence tomography. The RL-Deconv algorithm achieves results comparable to those in Ref. [16], however, it reduces the image contrast. ZS-DeconvNet improves contrast compared to the RL-Deconv algorithm but still exhibits sensitivity to thermal diffusion. In contrast, ZS-MSNN not only enhances contrast but effectively mitigates the thermal diffusion effect revealing the blood vessel network in Fig. 4(c).

To quantify the processing results, CNR was employed to compare different algorithms, as shown in Table II. S1-S4 denote samples in Fig. 4(a)-4(d), respectively. According to the evaluation results from the CNR, it is obvious that the proposed ZS-MSNN obtains the highest enhancement. Specifically, the improvement for S1-S4 was 202.60%, 105.93%, 95.32%, 116.62%, respectively. Therefore, the foregoing analysis validates the superior effectiveness of the proposed ZS-MSNN algorithm.

TABLE II
QUANTITATIVE EVALUATION BASED ON CNR.

| Method | Raw | RL Deconv | ZS-DeconvNet | ZS-MSNN |
|--------|------|-----------|--------------|---------|
| S1 | 4.62 | 5.25 | 7.12 | **13.98** |
| S2 | 5.73 | 9.27 | 9.44 | **11.80** |
| S3 | 4.49 | 6.12 | 8.67 | **8.77** |
| S4 | 10.95 | 12.86 | 12.88 | **23.72** |

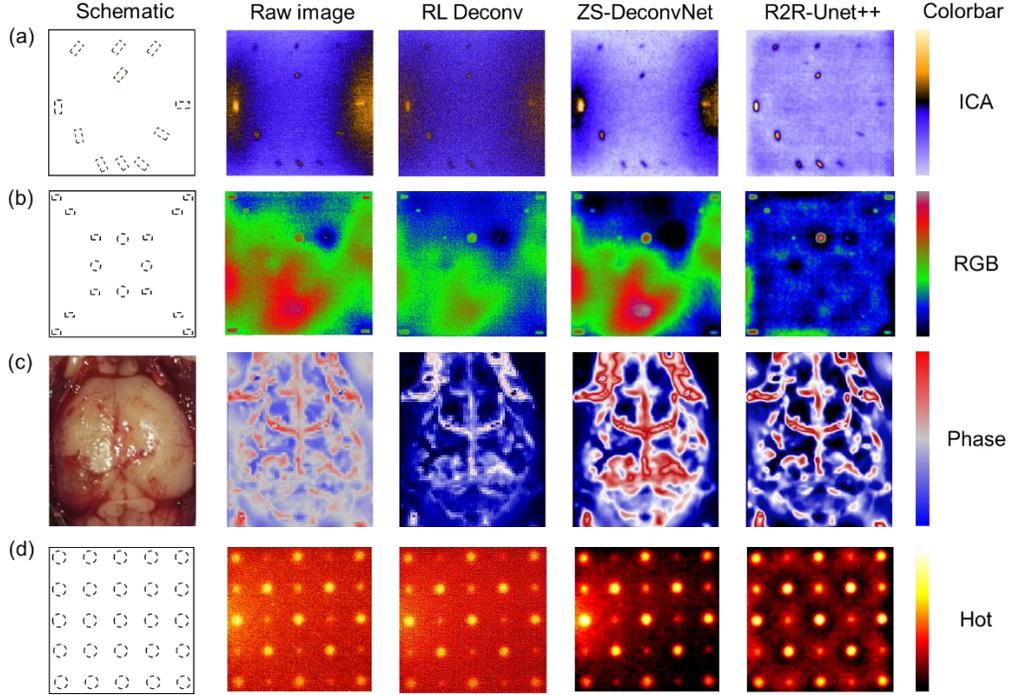

**Fig. 4.** The testing results for different datasets. (a) Plexiglass 1 with 11 rectangular defects. (b) Plexiglass 2 with 12 rectangular defects and 4 circular defects. (c) ex vivo mouse brain in Ref. [16]. (d) Plexiglass 3 with 25 circular defects.

*C. 3D Super-Resolution for Photothermal Coherence Tomography*

The original photothermal coherence tomography techniques can be traced back to truncated correlation tomography (TC-PCT), which is based on pulsed chirp excitation and match filtering. However, spectral aliasing caused by linear cross-correlation can significantly impede the performance of TC-PCT. The Mandelis group [8], [36], [37] proposed an enhanced truncated-correlation photothermal coherence tomography (eTC-PCT) technique that effectively addresses the spectral aliasing problem in photothermal tomographic imaging. The overall flowchart is shown in Fig. 5(a). In eTC-PCT the function generator produces a linear frequency modulation (LFM) pulsed chirp that controls the laser beam, *i.e.*, the in-phase reference truncated signal ($R_0$) and its quadrature ($R_{90}$) from the recorded excitation chirp [8]

$$R_0(t) = \sum_{m=0}^{p} \int_{m}^{m+W_T} \delta\left[t - \left(\frac{-\omega_1 + \sqrt{\omega_1^2 + 2\pi r(4m+1)}}{2r}\right) - W_T\right] dW_T \quad (15a)$$

and

$$R_{90}(t) = \sum_{m=0}^{p} \int_{m}^{m+W_T} \delta\left[t - \left(\frac{-\omega_1 + \sqrt{\omega_1^2 + 8m\pi r}}{2r}\right) - W_T\right] dW_T \quad (15b)$$

where $\omega_1$ is the starting angular frequency, $r = (\omega_2 - \omega_1)/T$ is the sweep rate, and $m = 0,1,2,\ldots p$, is the number of pulses to be generated. $\omega_2$ is the ending angular modulation frequency, and $T$ is the period of the LFM chirp. $\delta$ is the Dirac delta function. Then the cross-correlation (CC) technique is applied on the relaxation signals undergoing a time delay:

$$CC_{0,n}(t) = \int_{-\infty}^{\infty} R_{0,n}^*(t+\tau) T(\tau) d\tau \quad (16a)$$
$$CC_{90,n}(t) = \int_{-\infty}^{\infty} R_{90,n}^*(t+\tau) T(\tau) d\tau \quad (16b)$$

where $T$ is the relaxation signal, and $*$ is the complex conjugate operation. Finally, the cross-correlation amplitude and phase can be calculated from:

$$A_{CC,n} = \sqrt{CC_{0,n}^2 + CC_{90,n}^2} \quad (17a)$$
$$\emptyset_{CC,n} = \tan^{-1}(CC_{90,n}/CC_{0,n}) \quad (17b)$$

In the previous section, ZS-MSNN demonstrated promising denoising and deconvolution capabilities compared to conventional methods. To further highlight the effectiveness of ZS-MSNN, we applied this network to three-dimensional (3D) super-resolution imaging. While conventional methods in infrared thermography can extract useful information based on signal variation in the third dimension (time dimension), they often fall short in preserving 3D information. Statistical algorithms (*e.g.*, PCA) and frequency domain analysis (*e.g.*, PPT) can disrupt the original 3D data, and methods such as 2D Fourier filtering and general denoising algorithms offer limited improvement. Thus, the proposed ZS-MSNN is a valuable alternative.

The reference signals $R_0$ and $R_{90}$ are shown in Fig. 5(b). The delay time was set to 100 and the window size to 1. Because the delay time is limited by the heating time and the window size determines the depth resolution, the lower the window size, the higher the depth resolution [5]. The cross-correlation amplitude is shown in Fig. 5(c). This leads to an ultra-narrow pulse compression linewidth, such that the falling edge of the compressed pulse carries a highly depth-resolved signature of the photothermal features of the sample, the depth being coded in terms of delay time. Finally, the maximum photonic energy (amplitude peak) Eq. 17(a) was captured as the amplitude tomograms, and the location of the amplitude peak maximum in Eq. 17(b) was defined as phase tomograms (see Fig. 5(d)).

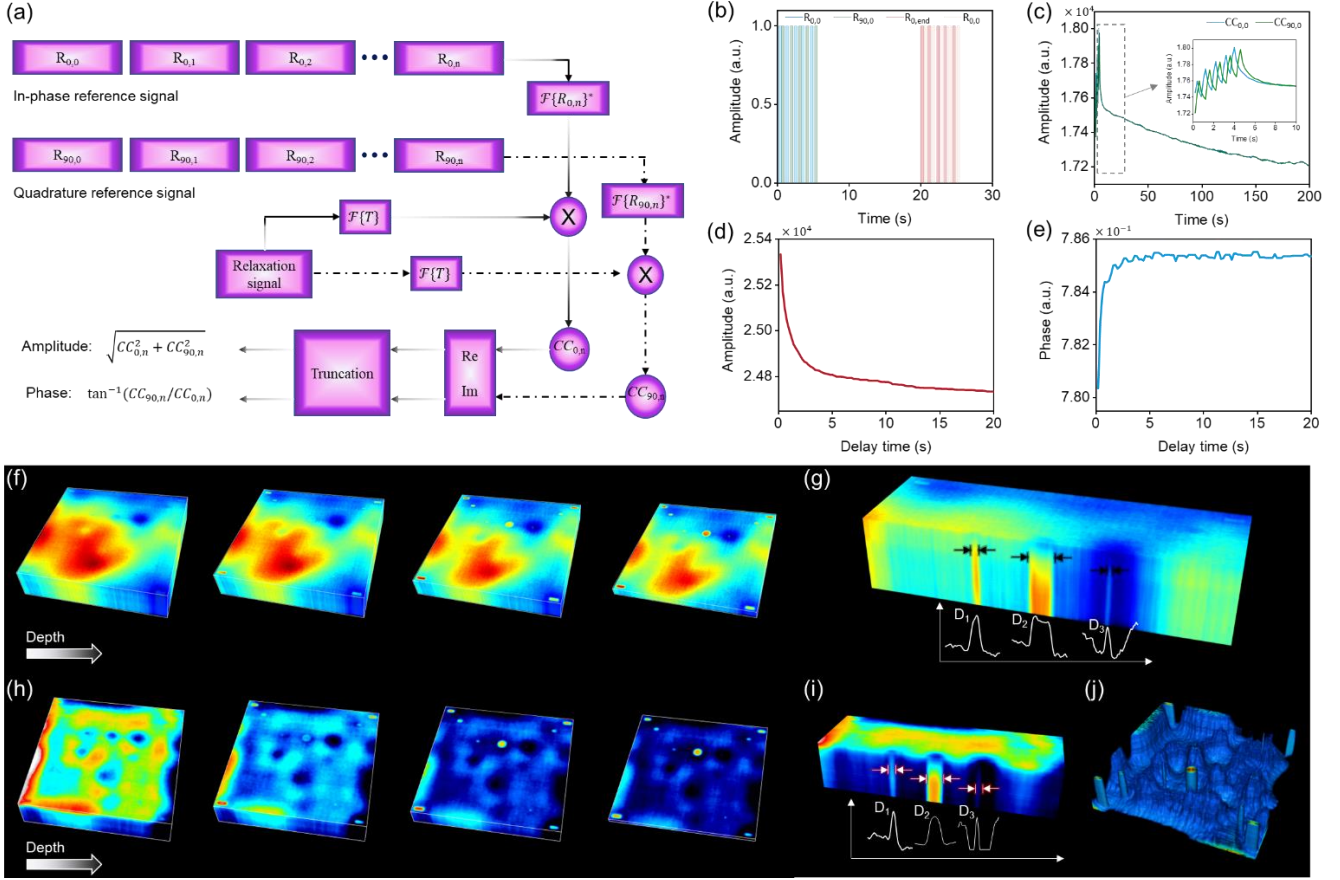

**Fig. 5.** 3D super-resolution for photothermal coherence tomography. (a) The schematic of eTC-PCT algorithm. (b) In-phase and quadrature reference signals. (c) Cross-correlation signals. (d) Cross-correlation amplitude results. (e) Cross-correlation phase results. (f) Tomograms of Plexiglass 2 at different depths. (g) Cross-section tomogram. (h) Tomograms processed by ZS-MSNN. (i) Cross-section tomogram processed by ZS-MSNN. (j) Tomogram with transparency and threshold optimized for defect profile inspection.

In eTC-PCT, amplitude tomograms offer higher depth resolution while phase tomograms provide better spatial resolution. Therefore, the amplitude tomograms were selected for monitoring the evolution of defects.

The original tomograms of the sample in Fig. 4(b) are shown in Fig. 5(f). Due to uneven heating and thermal diffusion, defects in the middle area of plexiglass 2 are not detectable. In addition, a "Red" area at the bottom covers all effective information. The cross-section tomogram is presented in Fig. 5(g). It is evident that the eTC-PCT algorithm approximately locks the lateral diffusion of the thermal wave along the depth direction. However, the temperature profile line is not a lateral Gaussian distribution, as described by the analytical solution of heat conduction $G(r) = e^{-kr}/(4\pi\alpha r)$ [38] where $r$ is the distance between the emission point on the defect and the observation point at the surface, $k$ is the familiar complex spatial wavenumber associated with a thermal wave, and $\alpha$ is the thermal diffusivity. This deviation is caused by spectral component leakage during the cross-correlation between reference and relaxation signals.

The ZS-MSNN was applied to process the results from the eTC-PCT algorithm. The tomograms at different depth slices are shown in Fig. 5(h). The defects, previously obscured by uneven heating and thermal diffusion in the middle area, are now observable. The cross-section tomogram is used to compare the lateral resolution of the original and processed images (see Fig. 5(i)), revealing that the temperature profile approximates a Gaussian distribution, particularly for defects $D_2$ and $D_3$. To further observe the defect profile, the original 3D tomogram was optimized by threshold segmentation, as shown in Fig. 5(j).

V. CONCLUSION

Uneven optical heating, accompanied by thermal diffusion, significantly limits the spatial and depth resolution of photothermal images. In this work, we constructed a real-time super-resolution imaging system, which is embedded with a novel zero-shot deep neural network, ZS-MSNN, to overcome the physical limitations of infrared thermography and photothermal coherence tomography. ZS-MSNN employs the U-Net++ as its backbone. This approach replaces the two outputs from the progressive two-stage U-Net used in previous work [26] with two outputs from a single DNN, thereby reducing the model's complexity and training time.

Unlike visual images, noise in thermal images coexists with effective information in the low frequency range, making it challenging to eliminate. To differentiate between noise and effective information in thermal images, we employed a Recorrupted-to-Recorrupted (R2R) scheme instead of the traditional Noise-to-Clean (N2C) scheme used in denoising DNNs for infrared thermography reported to-date. Experimental results demonstrated the superior performance of the R2R scheme. In addition, to enhance the denoising capability of ZS-MSNN, we applied high-pass Fourier filtering with a constant frequency component of 5 during pre-processing. Finally, to validate the generalization ability of the proposed model, we tested it on three plexiglass plates with various shapes and depths of defects, as well as an *ex vivo* mouse brain specimen from the literature [16]. The results from both 2D and 3D imaging show that ZS-MSNN exhibits excellent super-resolution imaging (denoising and deconvolution) capabilities.

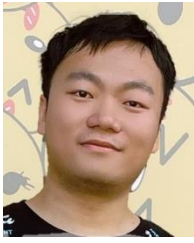

**Pengfei Zhu** received the B.Eng. degree in engineering mechanics from North University of China, Taiyuan, China, in 2019, and the M.Eng. degree in solid mechanics from Ningbo University, Ningbo, China, in 2022. He is currently working toward the Ph.D. degree in electrical engineering with Université Laval, Québec, Canada.

His research interests include non-destructive testing, infrared thermography, deep learning, terahertz time-domain spectroscopy, and photothermal coherence tomography.

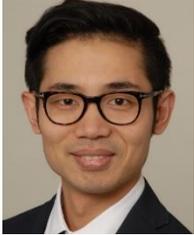

**Ziang Wei** received the M.Sc. degree in electrical engineering from RWTH Aachen University, Aachen, Germany, in 2017. He worked as an AI scientist at Fraunhofer IZFP and at the University of Applied Sciences in Saarbrüchen (htw saar), Germany, between 2017 and 2023. He is currently working toward the Ph.D. degree in electrical engineering with Université Laval, Québec, Canada.

His research interests include deep learning, data processing and Explainable AI.

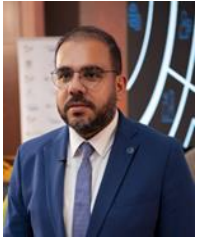

**Ahmad Osman** is a full professor at Saarland University of Applied Sciences in Germany and serves as Academic Chief at the Fraunhofer Institute for Nondestructive Testing. He holds dual PhD in Electrical Engineering from INSA-Lyon, France, and Computer Science, specializing in artificial intelligence, from Friedrich-Alexander University Erlangen-Nuremberg, Germany. His research focuses on the application of artificial intelligence in processing signals and images of data generated by nondestructive testing techniques, with the goal of advancing data interpretation and improving reliability in various engineering applications.

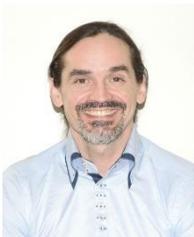

**Clemente Ibarra-Castanedo** received the M.Sc. degree in mechanical engineering (heat transfer) in 2000 from Université Laval, Quebec City, Canada, and the Ph.D. degree in electrical engineering (infrared thermography) in 2005 from the same institution. He is a professional researcher in the Computer Vision and Systems Laboratory at Université Laval and a member of the multipolar infrared vision Canada Research Chair (MIVIM). He has contributed to several research projects and publications in the field of infrared vision. His research interests are in signal processing and image analysis for the nondestructive characterization of materials by active/passive thermography, as well as near and short-wave infrared reflectography/transmittography imaging.

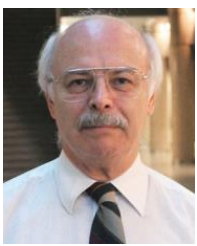

**Andreas Mandelis** FRSC, FCAE, FAPS, FSPIE, FAAAS, FASME, DF-IETI, PhD, is a full professor of Mechanical and Industrial Engineering; Electrical and Computer Engineering; and the Institute of Biomaterials and Biomedical Engineering, University of Toronto, and director of the Center for Advanced Diffusion-Wave and Photoacoustic Technologies at the University of Toronto. He is also the director of the Institute for Advanced Non-Destructive and Non-Invasive Technologies of the University. He has published more than 490 scientific papers in refereed journals and 190 proceedings papers in the fields of diffusion waves and photoacoustics. He has received numerous national and international prizes and awards and has several instrumentation and measurement methodology patents in photothermics, non-destructive evaluation, thermophotonics, optoelectronics, biophotoacoustics and new imaging methodologies.

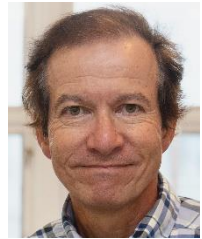

**Xavier Maldague** P.Eng., Ph.D. is full professor at the Department of Electrical and Computing Engineering, Université Laval, Québec City, Canada. He has trained over 50 graduate students (M.Sc. and Ph.D.) and contributed to over 400 publications. His research interests are in infrared thermography, NonDestructive Evaluation (NDE) techniques and vision / digital systems for industrial inspection. He is an honorary fellow of the Indian Society of Nondestructive Testing, fellow of the Canadian Engineering Institute, Canadian Institute for NonDestructive Evaluation, American Society of NonDestructive Testing. In 2019 he was bestowed a Doctor Honoris Causa in Infrared Thermography from University of Antwerp (Belguim).

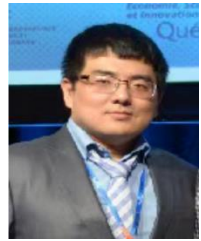

**Hai Zhang** is a full professor at Harbin Institute of Technology, Harbin, China. He received the Ph.D. degree in electrical engineering from Laval University, Quebec, QC, Canada, in 2017. He was a Postdoctoral Research Fellow with the Department of Mechanical and Industrial Engineering, University of Toronto, Toronto, ON, Canada. He was also a Visiting Researcher in Fraunhofer EZRT, Fraunhofer IZFP and Technical University of Munich, Germany. He has authored or coauthored more than 150 technical papers in peer-reviewed journals and international conferences.

He is also an Associate Editor for Infrared Physics and Technology, Measurement, and Quantitative InfraRed Thermography Journal. His research interests include nondestructive testing, industrial inspection, machine learning, medical imaging, infrared, and terahertz spectroscopy.